# Decrease of Entropy, Quantum Statistics and Possible Violation of Pauli Exclusion Principle

*Yi-Fang Chang*
*Department of Physics, Yunnan University, Kunming, 650091, China*
(e-mail: yifangchang1030@hotmail.com)

**Abstract**: We propose that a necessary condition of decrease of entropy in isolated system is existence of internal interactions. Then a theoretical development and some possible examples on decrease of entropy are researched. In quantum region, in particular for some Bose systems, decrease of entropy is more possible. The base of violation of Pauli exclusion principle is unified quantum statistics and attractive interactions, which corresponds also to decrease of entropy with internal interactions. Perhaps, some basis principles have essential relations.

## 1.Introduction

Recently, Pati [1] shown that in quantum Zeno effect setting the entropy of a quantum system decreases and goes to zero after a large number of measurements. He discussed the entropy change under continuous measurement model and show that entropy can decrease by using a more accurate measuring apparatus. Pattanayak, et al. [2], analyzed the behavior of the coarse-grained entropy decrease for classical probabilities and phase-space focusing in nonlinear Hamiltonian dynamics. Shokef, et al. [3], investigated a solvable model for energy-conserving nonequilibrium steady states. Two systems in contact do not reach the same effective temperature, and for this operational temperature the total entropy of the joint system can decrease. Erez, et al. [4], discussed thermodynamic control by frequent quantum measurements, and the corresponding entropy and temperature of both the system and the bath are found to either decrease or increase depending only on the rate of observation, and contrary to the standard thermodynamical rules that hold for memory-less (Markov) baths. Now the universality of the second law of the thermodynamics is querying from various regions.

We proposed that a necessary condition of decrease of entropy in isolated system is existence of internal interactions [5-7]. We discussed that since fluctuations can be magnified due to internal interactions under a certain condition, the equal-probability does not hold [7]. The entropy would be defined as

$$S(t) = -k \sum_r P_r(t) \ln P_r(t). \qquad (1)$$

From this or $S = k \ln \Omega$ in an internal condensed process, possible decrease of entropy is calculated.

The entropy of the composite system $A \oplus B$ verifies

$$S_q(A \oplus B) = S_q(A) + S_q(B) + (1-q)S_q(A)S_q(B). \qquad (2)$$

There will be $S_q(A \oplus B) < S_q(A) + S_q(B)$, if q>1. This is consistent with the system theory.

In fact, if various internal complex mechanism and interactions cannot be neglected, a state with smaller entropy (for example, self-organized structure) will be able to appear. In this case, the statistics and the second law of thermodynamics are possibly different [5,6]. Because internal interactions bring about inapplicability of the statistical independence, decrease of entropy in an isolated system is caused possibly.

## 2. Development of theory

We develop a generalized theory, in which entropy may increase or decrease, by using a similar method of the theory of dissipative structure. The total entropy in an isolated system is a formula [7]:

$$dS = dS^a + dS^i. \qquad (3)$$

Here $dS^a$ is an additive entropy, and $dS^i$ is an internal interacting entropy, whose rate assumes is

$$\frac{dS^i}{dt} = \sum_k J_k X_k, \qquad (4)$$

where $J_k$ are the flow, and $X_k$ are the corresponding internal forces. For the statistical independence, they are very small. If the flow

$$J_k = \sum_l L_{kl} X_l, \qquad (5)$$

so

$$\frac{dS^i}{dt} = \sum_k (\sum_l L_{kl} X_l) X_k. \qquad (6)$$

For the simplest case, $J_k = L_0 X_k$, so

$$\frac{dS^i}{dt} = \sum_k L_0 X_k^2. \qquad (7)$$

The rate of the interacting entropy is a nonlinear function of internal forces $X_k$.

In a real number field, there is $\frac{dS^i}{dt} < 0$ for $L_0 < 0$. This shows that some internal interactions may derive decrease of entropy. For $X_k = 0$ or $J_k = 0$, $dS^i = 0$, so $dS = dS^a > 0$. For a general case, $dS^i$ may be positive or negative. If the total entropy decrease,

$$dS = dS^a + dS^i_+ - dS^i_- < 0, \qquad (8)$$

the necessary condition will be

$$dS_-^i > dS^a + dS_+^i \geq dS^a. \qquad (9)$$

Further,

$$\frac{d^2 S^i}{dt^2} = \int (\sum_k J_k \frac{dX_k}{dt})dV + \int (\sum_k X_k \frac{dJ_k}{dt})dV. \qquad (10)$$

This way is completely analogous with the theory of dissipative structure [8].

## 3. Some possible examples of decrease of entropy

We researched the possibility of decrease of entropy in an isolated system for attractive process, internal energy, system entropy and nonlinear interactions, etc. [5-7].

In a deriving process of the fluctuation-dissipation theorem, there is [8]:

$$dS = (\frac{F}{T} - \overline{g}\alpha)d\alpha. \qquad (11)$$

When $X = \overline{g}\alpha > (F/T)$ and $d\alpha > 0$, i.e., for fluctuations magnified or for bigger force, dS<0.

For classically isolated gas, entropy is [8]:

$$S(T,V) = S_0 + nR\ln[(\frac{T}{T_0})^\alpha \frac{V}{V_0}]. \qquad (12)$$

Therefore, entropy should decrease when the volume becomes smaller or the temperature decreases.

The entropy of non-ideal gases is [9]:

$$S = S_{id} + N\log(1 - Nb/V). \qquad (13)$$

This is smaller than one of ideal gases, since b is four times volume of atom, b>0. It corresponds to the existence of interaction of the gas molecules, and average of forces between molecules is attractive [9]. The entropy of a solid is [9]:

$$S = 2\pi^2 VT^3 /15(\hbar\overline{u})^3, \qquad (14)$$

so dS<0 for dT<0. The free energy with the correlation part of plasma is [9]:

$$F = F_{id} - \frac{2e^3}{3}\frac{\pi^{1/2}}{(VT)^{1/2}}\left(\sum_a N_a z_a^2\right)^{3/2}. \qquad (15)$$

Correspondingly, the entropy is:

$$dS = S - S_{id} = -\frac{e^3}{3}\frac{\pi^{1/2}}{V^{1/2}}\left(\sum_a N_a z_a^2\right)^{3/2} T^{-3/2} < 0. \qquad (16)$$

There are electric attractive forces between plasma.

The second law of thermodynamics is based on neglect of fluctuations. But, fluctuations are important for light scattering, critical phenomena and phase transformation, etc., in which test for decrease of entropy is more possible.

In dielectric,

$$dU = TdS - p_0 dV + (V/4\pi)\varepsilon dD. \tag{17}$$

If dV=0, there will be dS<0 for $dU < (V/4\pi)\varepsilon dD$. In magnetic material,

$$dU = TdS - p_0 dV + (V/4\pi)\mu dB. \tag{18}$$

If dV=0, there will be dS<0 for $dU < (V/4\pi)\mu dB$. Or the entropy is [10]:

$$dS = du/T - (\mu_0 vM/C)dM. \tag{19}$$

If $du/T < (\mu_0 vM/C)dM$, there will be dS<0.

We discussed an attractive process based on a potential energy

$$U^i = -\frac{A}{r}, \tag{20}$$

in which entropy decreases [7]. Entropy of the ideal gases is

$$S = C_V \ln T + nR \ln V + S_0. \tag{21}$$

For an equal-temperature process T=constant,

$$dS = S_f - S_i = nR \ln(V_f / V_i\}. \tag{22}$$

This decreases dS<0, when $V_f < V_i$, i.e., for the attractive process. Conversely, the entropy increases dS>0, when $V_f > V_i$.

We found that negative temperature derives necessarily decrease of entropy [11]. Negative temperature is based on the Kelvin scale and the condition dU>0 and dS<0. Conversely, there is also negative temperature for dU<0 and dS>0. But, negative temperature is contradiction with usual meaning of temperature and with some basic concepts of physics and mathematics. It is a notable question in nonequilibrium thermodynamics.

**4. Quantum statistics**

This is known that these are different quantum statistics, Fermi-Dirac (FD) statistics, Bose-Einstein (BE) statistics, and Maxwell-Boltzmann (MB) statistics. The entropy of Fermi gas and entropy of Bose gas are different. In a special issue <Entropy> (March 2004) edited by Nikulov, et al., many aspects for quantum limits to the second law of thermodynamics are discussed [12].

In quantum statistics the free energy corrected to MB value is [9]:

$$F = F_{MB} \pm \frac{\pi^{3/2}}{2g} \frac{N^2 \hbar^3}{VT^{1/2} m^{3/2}}, \tag{23}$$

where and later the upper sign corresponds to FD statistics and the lower sign corresponds to BE statistics. Correspondingly, the entropy is:

$$S = -\left(\frac{\partial F}{\partial T}\right)_V = S_{MB} \pm \frac{\pi^{3/2}}{2g} \frac{N^2 \hbar^3}{VT^{3/2} m^{3/2}}. \qquad (24)$$

Therefore, $dS = S - S_{MB} > 0$ for Fermi gases, and $dS < 0$ for Bose gases, i.e., the entropy in BE statistics is smaller than in MB statistics. In FD statistics the quantum exchange effects lead to the occurrence of an additional effective repulsion between the particles, and in BE statistics there is an effective attraction between the particles [9]. This is consistent with decrease of entropy for attractive process [7].

The second virial coefficient may be calculated with allowance for the quantization of the binary interaction of the gas particles, and the atoms obey BE statistics. The free energy is [9]:

$$F = F_{id} - \frac{8N^2}{V}\left(\frac{\pi \hbar^2}{m}\right)^{3/2} Z_{int} T^{-1/2}. \qquad (25)$$

Correspondingly, the entropy is:

$$dS = S - S_{id} = -\frac{4N^2}{V}\left(\frac{\pi \hbar^2}{m}\right)^{3/2} Z_{int} T^{-3/2}. \qquad (26)$$

This is smaller than entropy of ideal gases.

It is well known that entropy is measurement of disorder in a system. In phase transformation the crystallization of an over-cooling liquid or of a supersaturated solution is surely an ordering process.

The cooling principle is entropy reduced. The difference of entropies between the normal state and the superconductive state is:

$$dS = S_s - S_n = \frac{H_c V}{4\pi} \frac{\partial H_c}{\partial T}, \qquad (27)$$

where $S_s$ is entropy of the normal state, and $S_s$ is one of the superconductive state. Such $dS < 0$ for $\frac{\partial H_c}{\partial T} < 0$. A superconducting state is more order than a normal state. This phase transformation from the normal state to the superconducting state is a condensation process. Generally, any condensation process, in which attractive interactions exist, should be one of decrease of entropy [7].

In the Ginzburg-Landau theory the difference in the free energies of the superconducting and normal states:

$$F_s - F_n = -V(a^2/2B)(T_c - T)^2. \qquad (28)$$

Correspondingly, the entropy is:

$$dS = S_s - S_n = -V(a^2/B)(T_c - T). \qquad (29)$$

Such $dS < 0$ for $T < T_c$.

The occurrence of superfluidity in a Fermi system is due to the Cooper effect, the formation of bound states (pairing) by mutually attracting particles. We predict that entropy will decrease for the Cooper effect, further, for the Bose-Einstein condensation, for superconductivity, and superfluidity.

The superfluid helium and its fountain effect must suppose that the helium does not carry entropy, so that the second law of thermodynamics is not violated [8]. It shows that superfluidity will possess zero-entropy, but this cannot hold because zero-entropy corresponds to absolute zero according to the third law of thermodynamics. For the liquid or solid $He^3$ the entropy difference [8] is $\Delta S = S_l - S_s >0$ (for higher temperature), =0 (for T=0.3K), <0 (for lower temperature). Such a solid state with higher entropy should be disorder than a liquid state in lower temperature!

Of course, the above examples are not already under the thermodynamic equilibrium condition. In the evolutionary process and the phase transformation, the systems cannot be in thermal equilibrium; this is true for various systems in biology and society.

When there is a critical point, a continuous transformation can be effected between any two states of the substance [9]. Interactions between the molecules are different for the two states, and internal symmetries are different, so entropies are also different. In the unsymmetrical phase, the entropy is [9]:

$$S = S_0 + (a^2/2B)(T - T_c). \tag{30}$$

At the transformation point itself, this expression becomes $S_0$. For T>$T_c$, $dS = S - S_0 >0$; but for T<$T_c$, $dS = S - S_0 <0$. A symmetrical state is more order than unsymmetrical state. At a critical point of phase transformation of the second kind, there is [6]:

$$dS = S - S_0 = -a\eta^2 < 0 \quad \text{(for } a>0\text{)}, \tag{31}$$

where $\eta$ is an order parameter.

These results are similar in the Landau-Devonshire theory and the Landau-de Gennes theory of phase transformation.

Eliezer and Weiner introduced the temperature concept associated with the energy of a particle reaction into the thermodynamics of field theory. Here the entropy density is [13]:

$$S_1 = -(\partial \widetilde{V}_{eff}/\partial T) = 4N_0 T^3 \quad \text{for } T > T_c, \tag{32}$$

$$S_2 = 4N_0 T^3 - \lambda N^4 (T_c^2 - T^2)T \quad \text{for } T \leq T_c. \tag{33}$$

Such $dS = S_2 - S_1 = -\lambda N^4 (T_c^2 - T^2)T < 0$ This temperature cannot also be infinite. For weak and electromagnetic interactions, this implies that the phase transformation predicted within a unified gauge theory of weak and electromagnetic interactions should be looked for in the particle reaction at very high energy. In the phase transformation hadronic matter has superfluid properties. Further, we extended this phase transformation theory of hadron, and proposed a many shell-state model of the particle-structure, in which various phase transformations among particles, quarks, subquarks, preprons, and sandons will appear with increase of energy, and suggested the symmetry-statistics duality of particle [14]. The second law of the thermodynamics is violated by quantum measurements [4].

Although in an isolated system the gravitational field cannot be screened completely, but the electromagnetic field may be screened completely. It is very easily that strong and weak interactions as short-distance ones are screened.

In a system with internal interactions, the fluctuation can be magnified, for example, in those processes of phase transformation. When the order parameter of a system comes to a threshold value, a phase transformation occurs, and self-organization will take place, as in synergetics [15]. Simultaneously, the entropy will decrease continuously, and a final state with lower entropy will be reached. In this case various microscopic states are not equally probable. If entropy is seen as a degree of freedom, the interaction will reduce the degree of freedom.

In phase transformation, structures and symmetries are different, corresponding internal energies and entropies are also different. At a critical point of phase transformation, entropy can increase or decrease, i.e., possesses two-direction property. This corresponds to reversibility of transformation between order and disorder. Of course, the general phase transformation is an open system. But, if input energy is interrupted at the critical point, it will become an isolated system.

The specific heat $C_V = TdS/T$ is anomalous, i.e., entropy $S = \int_0^T (C_V/T)dT$ is anomalous. This implies decrease of entropy after our analysis [5-7].

A nucleation in phase transformation for solids, which may be formed due to thermal fluctuation, is an ordered process. The second law of thermodynamics is corrected, it will be conducive to develop freely theories on nucleation and grow in dynamics of phase transformations.

The chemical reactions are very complex, and include oscillation, condensation, catalyst and self-organization, etc. In these case changes of entropy may increase or decrease. The second law of thermodynamics is based on an isolated system and statistical independence [9]. If fluctuations magnified due to internal interactions exist in the system, entropy will decrease possibly [5-7]. In chemical reactions there are various internal interactions, so some ordering processes with decrease of entropy are discussed [16].

Under some conditions usual entropy may decrease. It is more possible that the structural entropy, the topological entropy, the metric entropy, and information entropy, etc., various entropies decrease, I think. In an isolated system, increase of information should be obvious. For example, new idea can be produced continuously from interchange of thinking.

## 5. Decrease of entropy and possible violation of Pauli exclusion principle

The present experimental tests have proved high precisely the validity of the Pauli exclusion principle (PEP) in usual cases. But, based on some experiments and theories of particles at high energy, we suggested that particles at high energy possess a new statistics unifying BE and FD statistics [17], for example, a possible unified distribution is:

$$y_\Gamma = \frac{\beta^\alpha}{\Gamma(\alpha)} x^{\alpha-1} e^{-\beta x}. \qquad (34)$$

This agrees quantitatively with scaling, the multiplicity, and large transverse momentum, etc. Such PEP would not hold at high energy [17,18]. Santilli was the first to proposed the rest of PEP, in particular, under strong interactions [19,20]. Ignatiev-Kuzmin and Greenberg-Mohapatra [21] proposed that PEP has a small violation for any natural substance, which should contain a fraction of order $\beta^2$ of anomalous atoms and nucleons. The future experiments should be combined widely with various theories of hidden and obvious violation of PEP [22]. We proposed some possible tests of the violation of PEP [18,14,22], for example, the excited high-n atoms, the various nuclei at high energy, dineutrons in extremely neutron-rich nuclei, and gamma ray sources in high energy astrophysics, etc.

In fact, the parastatistics, the fractional statistics [23], anyon, and the fractional quantum Hall effect, etc., have some contradictions with the standard theory in which two types of different particles and their properties are distinguished from the spin-statistics stringently. Even in the nonabelian gauge field theory there is the ghost particle whose spin is zero, but which agrees with anticommutation relation. They correlate to various theories relevant to possible violation of PEP, including some obvious and hidden ones [24,22].

Mohapatra [25] predicted the presence of a neutral spin-3/2 hadron with mass in the 1-2 GeV range by using infinite statistics. It implies the violation of PEP at 1-2 GeV. According to the uncertainty principle we expected that usual high energy is about 2-20 GeV for particles [17]. In different regions, for instance, nuclei, multiplicity and celestial body, etc., there should be corresponding threshold values for high energy.

The possible violation of PEP corresponds to development of statistics and unified statistics [17]. On the other hand, possible decrease of entropy in an isolated system must have some internal interactions, in particular, for attractive process [5-7]. The basis of thermodynamics is the statistics, in which various interactions among the subsystems should not be considered [9]. The base of PEP is FD statistics, whose condition is also that interactions among particles are not considered. PEP has repulsion, and its violation should be attraction, for example, Cooper pair. Both aspects are consistent. Generally, both are contacted with interactions and nonlinearity. If there are attractive forces in some systems, entropy will decrease, and PEP will be violated possibly, and FD and BE statistics will be unified. Both are related with black hole [18] and white hole. Perhaps, they have essential relation. Their developments may be analogous each other. Further, the violation of PEP and decrease of entropy should be relevant to nonlinearity [26,7].

In a word, some basic principles are possibly correlated each other. For example, the high energy objects may be applied to test PEP and PCT invariance, etc. [27,22].

**References**
1. Pati, A.K., 2000. arXiv.org/quant-ph/0006089.
2. Pattanayak, A.K., Brooks, D.W.C., de la Fuente, A., et al., 2005. Phys.Rev. A72, 013406.
3. Shokef, Y., Shulkind, G. & Levine, D., 2007. Phys.Rev. E76, 030101(R).
4. Erez, N., Gordon, G., Nest, M. & Kurizki, G., 2008. Nature. 452, 724.
5. Chang Yi-Fang, In Entropy, Information and Intersecting Science. Yu C.Z., Ed. Yunnan Univ. Press. 1994. p53-60.
6. Chang Yi-Fang. 1997. Apeiron. 1997, 4,4,97.
7. Chang Yi-Fang, 2005. Entropy. 7,3,190.
8. Reichl, L.E., A Modern Course in Statistical Physics. Univ.of Texas Press. 1980.

9.Landau, L.D. & Lifshitz, E.M., Statistical Physics. Pergamon Press. 1980.

10.Holman, J.P., Thermodynamics. Third Edition. McGraw-Hill. 1980.

11.Chang Yi-Fang, 2007. arXiv:0704.1997.

12.Nikulov, A.V. & Sheehan, D.P. (Ed.), 2004. Entropy. 2004, 6,1,1-232.

13.Eliezer, S. & Weiner, R.M., 1976. Phys.Rev. D13,1,87.

14.Chang Yi-Fang, New Research of Particle Physics and Relativity. Yunnan Science and Technology Press. 1989; 1990. Phys.Abst. 93:No1371.

15.Haken, H., Synergetics. An introduction. Springer-Verlag. 1977.

16.Chang Yi-Fang, 2008. arXiv:0807.0256.

17.Chang Yi-Fang, 1984. Hadronic J. **7**,5,1118.

18.Chang Yi-Fang, 1984. Hadronic J. 7,6,1469.

19.Santilli, R.M., 1978. Hadronic J.1,223; 574.

20.Ktorides, C.N., Myung,H.C. & Santilli, R.M., 1980. Phys.Rev. D22,892.

21.Greenberg, O.W. & Mohapatra,R.N., 1987. Phys.Rev.Lett. 59,2507;1989. Phys.Rev. D39,2032.

22.Chang Yi-Fang, 1999. Hadronic J. 22,3,257.

23.Smerzi,A., 1996. Phys.Rev.Lett.76,559.

24.Chang Yi-Fang, 1992. Hadronic Mechanics and Nonpotential Interactions (Myung,H.C.ed.), Par2. Nova Science Publishers,Inc. p169-175.

25.Mahapatra,R.N., 1990. Phys.Lett. B242,407.

26.Chang Yi-Fang, 1991. Proc.of the 4th Asia-Pacific Phys.Conf. V2. World Scientific. p1483-86.

27.Chang Yi-Fang, 1999. J. New Energy. 4,1,31.